\documentclass[twocolumn]{aastex631}
\usepackage{booktabs}
\usepackage{dcolumn}
\newcolumntype{d}[1]{D{.}{.}{#1}}
\usepackage{multirow} 
\usepackage{amsmath}
\usepackage{graphicx,verbatim}
\graphicspath{{figures/}}

\begin{document}

%\title{Detecting Dark Matter by Gravitational-Wave Background from Early-Stage Extreme Mass-Ratio Inspirals in the Milky Way Center}
\title{Probing Dark Matter Spike with Gravitational Waves from Early EMRIs in the Milky Way Center}

\author{Chen Feng}
%\email{fengchen23@mails.ucas.ac.cn}
\affiliation{School of Astronomy and Space Science, University of Chinese Academy of Sciences (UCAS), Beijing 100049, China}

\author{Yong Tang}
%\email{tangy@ucas.ac.cn}
\affiliation{School of Astronomy and Space Science, University of Chinese Academy of Sciences (UCAS), Beijing 100049, China}
\affiliation{School of Fundamental Physics and Mathematical Sciences, Hangzhou Institute for Advanced Study, UCAS, Hangzhou 310024, China}
\affiliation{International Centre for Theoretical Physics Asia-Pacific, UCAS, Beijing 100190, China}

\author{Yue-Liang Wu}
\affiliation{School of Fundamental Physics and Mathematical Sciences, Hangzhou Institute for Advanced Study, UCAS, Hangzhou 310024, China}
\affiliation{International Centre for Theoretical Physics Asia-Pacific, UCAS, Beijing 100190, China}
\affiliation{Institute of Theoretical Physics, Chinese Academy of Sciences, Beijing 100190, China}
 
\begin{abstract}
Cold dark matter may form dense structures around supermassive black holes (SMBHs), significantly influencing their local environments. These dense regions are ideal sites for the formation of extreme mass-ratio inspirals (EMRIs), in which stellar-mass compact objects gradually spiral into SMBH, emitting gravitational waves (GWs). Space-based gravitational-wave (GW) observatories such as LISA and Taiji will be sensitive to these signals, including early-stage EMRIs (E-EMRIs) that persist in the low-frequency band for extended periods. In this work, we investigate the impact of dark matter-induced dynamical friction on E-EMRIs in the Milky Way Center, model its effect on the trajectory, and calculate the resulting modifications to the GW spectrum. Our analysis suggests that this influence might be sizable and lead to detectable deviations in the spectrum, namely suppression at low frequencies and enhancement at high frequencies, therefore providing a potential probe for dark matter with future GW detectors in space, such as LISA and Taiji.
\end{abstract}


\section{Introduction}\label{sec:intro}
Observational evidence from galactic rotation curves \citep{1939dmrotation}, gravitational lensing in galaxy clusters \citep{Dell'Antonio_1996Len}, and the Bullet Cluster \citep{Clowe_2006Bulletcluster} strongly supports the existence of dark matter. However, the spatial distribution of dark matter in galaxy center has not yet been clearly determined~\citep{dmprofile, Ullio_2001}. Theoretical studies suggest that the presence of an adiabatically growing supermassive black hole (SMBH) in the galactic core can significantly enhance the density of the surrounding cold dark matter~\citep{spike,  spikerelativistic, Lacroix:2013qka, spikeFerrer_2017, Zhang:2024hrq, Zhang:2025mdl}. The intense gravitational field of the SMBH steepens the density profile, leading to the formation of a dark matter spike in the inner regions. This phenomenon implies a strong concentration of dark matter near the SMBH, resulting in an exceptionally dense environment.

The density environments surrounding SMBH facilitate diverse astrophysical processes capable of generating gravitational waves (GWs). As a fundamental observational probe in astrophysics, GWs would provide unprecedented insights into cosmic phenomena and compact objects \citep{Bian:2025ifp}. The development of space-based GW observatories, particularly the Laser Interferometer Space Antenna (LISA)~\citep{amaroseoane2017laserinterferometerspaceantenna} and Taiji \citep{Hu:2017mde}, will facilitate the detection of millihertz GWs in the coming decades. This observational capability promises advances in understanding the dynamics of SMBHs located in galactic centers and their interactions with the surrounding environment. Among various potential sources, the extreme mass ratio inspirals (EMRIs) are particularly significant, describing the gradual inspiral of compact objects (COs) into SMBHs \citep{Amaro_Seoane_2007, Amaro_Seoane_2019}. During this process, a CO follows a decaying orbit, with the orbital decay accelerating as it spirals inward. Ultimately, the system undergoes a rapid plunge, culminating in the merger of CO with the SMBH and the emission of intense gravitational-wave (GW) signals \citep{Peters1964, 2020HandbookOG}.

While the CO inspirals toward the SMBH eventually, objects at large separations evolve slowly and remain in the early inspiral phase for extended periods~\citep{Peters1964}. These long-lived systems are classified as early EMRIs (E-EMRIs)~\citep{seoane2024eemris, Seoane:2025gfh}. Although the intrinsic occurrence rate of EMRIs is relatively low, their extended residence within the GW observation band renders them ideal targets for high-sensitivity detectors such as LISA and Taiji \citep{colpi2024lisadefinitionstudyreport, Barak2004, Barak2007, Babak_2017, Amaro_Seoane_2015, Klein_2016}. The persistence of their signals not only offers invaluable insight into SMBHs and their environments \citep{Baryakhtar:2022hbu, Yue:2024xhf, Karydas:2024fcn, Yue:2018vtk}, but also provides a unique window into the physical mechanisms that govern GW sources \citep{Cui:2025bgu}. Since the inspiral of COs occurs near the SMBH, the formation and evolution of EMRIs are influenced by environmental factors in its vicinity \citep{Fu:2024cfk, Zhang:2024ogc, Shen:2025svs, Zhang:2024ugv}. We consider this environment to be primarily composed of COs, accretion flow \citep{science1240755}, and dark matter \citep{Navarro_1996, Planck2018, cirelli2024darkmatter}, with dark matter as one of the key contributors. 

In this work, we examine the process of relaxation-driven EMRIs formation and find that dark matter-induced dynamical friction promotes orbital circularization, thereby suppressing EMRIs formation by hindering relaxation-driven eccentricity growth. We calculate the number of EMRIs in the Galactic Center and demonstrate that dynamical friction reduces the overall source population. This effect suppresses the GW background generated by unresolved E-EMRIs, particularly in the low-frequency regime, through accelerating the orbital evolution. Our findings suggest that it is possible to probe dark matter spikes using future space-based GW detectors.

This paper is organized as follows. In Sec.~\ref{sec:formation} we describe the EMRIs formation scenario, including the role of dark matter in this process. Then in Sec.~\ref{sec:number of source} we estimate source populations in the Galactic Center and their modulation by dark matter. And in Sec.~\ref{sec:background} we calculate the E-EMRIs GW spectrum, quantifying dark matter's effect on the signal strain. Finally in Sec.~\ref{sec:conclusion} we summarize the results.

\section{EMRIs formation}\label{sec:formation}
We shall first introduce relaxation-driven EMRIs formation and establish the terminology. After presenting the formation scenario, we compare the various environmental components of the Galactic Center and demonstrate the dominance of dark matter. We then analyze dark matter dynamical friction and its impact on EMRIs formation.

\subsection{Relaxation Process}\label{subsec:encounter}

While a test object (e.g., black hole, neutron star, or white dwarf) moves through the environment consisting of field objects around the SMBH, its orbit can be approximated as a Keplerian orbit. If the field objects consist of COs whose mass is non-negligible compared to that of the test object, gravitational encounters can lead to relaxation. This relaxation process may drive the orbit to become highly eccentric. As a result, the periapsis can approach sufficiently close to the SMBH, triggering the emission of GWs and subsequent orbital decay \citep{Amaro_Seoane_2018}.

The characteristic timescale for the relaxation process to change the eccentricity $e$ is denoted as the relaxation time $t_{\rm{rlx}}$, and is given by \citep{Spitzer2016-xf}
\begin{equation}
    t_{\rm{rlx}} = \frac{0.34 \sigma_{\rm{f}}^3}{G^2 m \rho \ln \Lambda}(1 - e),
\end{equation}
where $\sigma_{\rm{f}}$ is the velocity dispersion of field stars, $m$ is their mass, $\rho$ is the local density, and $\ln \Lambda$ is the Coulomb logarithm.
%A more detailed discussion can be found in Appendix~\ref{appendix:A}. 

While the relaxation process drives the orbit to become highly eccentric, GW emission tends to circularize the orbit. A test object, initially located far from the SMBH, undergoes relaxation, causing its periapsis to shrink sufficiently close to the SMBH for GW emission to become significant. This leads to orbital decay and ultimately results in the object plunging into the SMBH.

The timescale for GW emission to alter orbital eccentricity is defined as the gravitational time, expressed as
\begin{equation}
    t_{\rm{gw}} = \frac{1 - e}{[d(1 - e)/dt]_{\rm{gw}}} = -\frac{1 - e}{(de/dt)_{\rm{gw}}}.
\end{equation}
Ignoring the effects of other background field objects, a system can form an EMRI if the timescale for GW emission is shorter than the relaxation time. In addition, the orbit resulting from relaxation must avoid a direct plunge into the SMBH \citep{Amaro_Seoane_2018}. These conditions can be expressed as
\begin{equation}
    \begin{aligned}
        t_{\rm{gw}} &< t_{\rm{rlx}}, \\
        a(1 - e) &> \frac{8GM}{c^2} \, \mathcal{W},
    \end{aligned}
\end{equation}
where the quantity $8GM/c^2$ is referred to as the plunge radius \citep{Will_2012}. The factor $\mathcal{W}$ accounts for the influence of orbital asymmetry on the location of the last stable orbit in the Kerr and Schwarzschild cases \citep{Amaro_Seoane_2013}, and is taken to be 0.26 in this work. By applying these conditions, we can identify a critical semi-major axis $a_{\rm{cri}}$, which determines the maximum value of $a$ at which a system can form an EMRI.

\subsection{Envrionment Around SMBH}\label{subsec:env}
The SMBH provides an extremely dense environment. In its vicinity, in addition to COs, there is also a significant distribution of dark matter and accretion flow. An SMBH at the galactic center can redistribute the surrounding dark matter, leading to the formation of a density spike. This process can be modeled by the adiabatic growth of the black hole. Initially, the dark matter halo near the galactic center follows a power-law density profile~\citep{Quinlan_1995, Navarro_1996}. After the adiabatic growth of the black hole, the profile remains a power law but becomes significantly steeper. The resulting spike density profile is given by~\citep{spike}
\begin{align}\label{eq:dmprofile}
    \rho(r) = \rho_{\rm{c}} \left( \frac{r_{\rm{c}}}{r} \right)^{\gamma},\, \gamma = \frac{9-2\beta}{4-\beta},
\end{align}
where $\beta$ is the halo power law index and $\gamma$ is the spike power-law index. $r_{\rm{c}}$ denotes the characteristic radius of the system, and $\rho_{\rm{c}}$ is the corresponding density at this radius. The exact value of the spike power-law index $\gamma$ remains uncertain and depends on the property of dark matter~\citep{Xie:2025udx, Zhang:2025mdl}, formation history and dynamical evolution of the Galactic Center. In this study, we take a phenomenological approach and consider representative values of $\gamma = 2,\; 2.5,\; 3$ and $3.5$ to explore its potential impact.

Based on observations of the orbital dynamics of the S2 star, \cite{2024constarin} established an upper limit for the total enclosed mass within the orbital radius of S2 to be approximately $1200\,M_{\odot}$. As we shall show later in Sect.~\ref{subsec:formation}, EMRIs are approximately formed at $0.01\,\rm{pc}$ from the SMBH. Hence we set $r_{\rm{c}} = 0.01\,\rm{pc}$ for convenience. For different $\gamma$, we can determine the corresponding upper limit on $\rho_{\rm{c}}$. When we choose $\gamma$ to be 2, 2.5, 3, and 3.5, the upper limits are $\rho_{\rm{c}} \lesssim 1.1 \times 10^8$, $6.5 \times 10^7$, $3.4 \times 10^7$, and $1.5 \times 10^7 \, M_{\odot}/\mathrm{pc}^3$, respectively.

Sgr A* is classified as an extremely low-luminosity source. Its broadband spectral energy distribution is most accurately modeled by radiatively inefficient accretion flows, characterized by a power-law density profile expressed as $n \propto r^{-\alpha}$. The L\textquotesingle-band emission from the S2 star imposes constraints on the accretion flow density profile in the vicinity of Sgr A*. In the ambient shock scenario, these constraints establish an upper limit on the slope of the power law of $3.2$ and on the density of ambient numbers, which is restricted to below $1.87 \times 10^9 \,\rm{cm}^{-3}$ near the periapse of S2 \citep{Hosseini2020}.
\begin{figure}[t]
    \centering
    \includegraphics[width=1\linewidth]{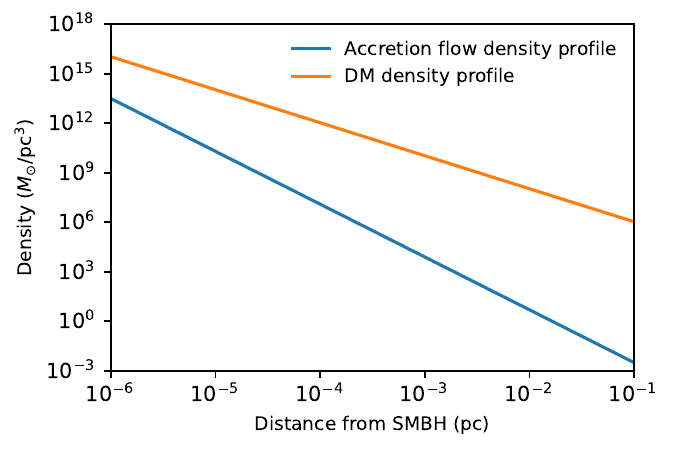}
    \caption{The comparison of the density profiles between dark matter and the accretion flow. Here, the dark matter power-law index is $\gamma = 2$. In contrast, the accretion flow follows a power-law index of $\alpha = 3.2$, with a number density of $1.87 \times 10^9 \,\rm{cm}^{-3}$ at the periapse of S2.}
    \label{fig:densityprofile}
\end{figure}

In Fig.~\ref{fig:densityprofile} we plot the densities of dark matter with $\gamma=2$ and accretion flow. We can see the density of dark matter is significantly higher than that of the accretion flow at all distances of our interested region. As a result, we consider only dark matter and the relaxation effect induced by COs in this study. 

\subsection{Dynamical friction of dark matter spike}\label{subsec:DM fric}
Since we consider dark matter as microscopic particles whose mass is negligible compared to that of the CO, dark matter can exert a dynamical friction on its orbital motion. %A more detailed discussion is provided in Appendix~\ref{appendix:A}.  
Numerous investigations have examined dynamical friction across diverse dark matter scenarios, including ultra-light dark matter \citep{Traykova:2021dua, Gorkavenko_2024, Wang_2021, Boey_2024, Vicente_2022, Duque:2023seg, Aurrekoetxea:2023jwk} and self-interacting dark matter \citep{Fischer_2024, Glennon_2023, Koo_2025, Boudon_2023, Sabarish:2025hwb}, among others \citep{Zhang:2022rfr, Berezhiani_2023, Zhou_2024, Shen:2023kkm, Mitra:2025tag, Yang:2024cnd, Feng:2024oab, Liang:2022gdk, Chan:2022gqd}. In this work, we illustrate with the cold dark matter scenario in which a spike can form around an SMBH. 
 
The acceleration $a_{\rm{df}}$ induced by dynamical friction from dark matter spike is given by~\citep{2024df}
\begin{multline}
    \overline{a}_{\rm{{df}}} = 
    - \frac{16 \pi^2 G^2 m\rho}{v^2}
    \int_0^{v_{\rm{esc}}} dv_{\rm{f}} \, v_{\rm{f}}^2 f(v_{\rm{f}}) \mathcal{H}(v, v_{\rm{f}}, p_{\rm{max}}).
\end{multline}
Here, $v_{\rm{esc}}$ is the escape velocity and $\mathcal{H}$
\begin{align}
    \mathcal{H}(v, v_{\rm{f}}, p_{\rm{max}}) \approx 
    \begin{cases} 
        \ln \Lambda & \rm{if }\; v > v_{\rm{f}}, \\[10pt]
        \ln \left( \frac{v_{\rm{f}} + v}{v_{\rm{f}} - v} \right) - 2 \frac{v}{v_{\rm{f}}} & \rm{if }\; v < v_{\rm{f}}.
    \end{cases}
\end{align}
And the Coulomb logarithm $\Lambda$ is given by
\begin{align}
    \Lambda = \frac{p_{\rm{max}}v_{\rm{c}}^2}{Gm},
\end{align}
where $p_{\rm{max}}$ denotes the maximum impact parameter, typically defined as the characteristic scale of the system. In this context, we set $p_{\rm{max}}$ to 1 pc, which corresponds to the influence radius of the SMBH. $v_{\rm{c}}$ represents the circular velocity in the vicinity of the SMBH.

Then the time-evolution equations of the orbit are
\begin{multline}
    \left\langle \frac{da}{dt} \right\rangle_{\rm{df}} =
    \frac{(1 - e^2)^2}{\pi k^3 a^2}
    \int_{0}^{2\pi} 
    \frac{(1 + e \cos \theta)^{-2} \epsilon(r, v)}{(1 + e^2 + 2e \cos \theta)^{1/2}} \, d\theta, 
\end{multline}
\begin{multline}
    \left\langle \frac{de}{dt} \right\rangle_{\rm{df}} =
    \frac{(1 - e^2)^3}{\pi k^3 a^3} \times \\
    \int_{0}^{2\pi}  
    \frac{(e + \cos \theta) \epsilon(r, v)}{(1 + e^2 + 2e \cos \theta)^{3/2} (1 + e \cos \theta)^2} \, d\theta.
\end{multline}
Here, $k$ is the orbital angular frequency and $\theta$ is the true anomaly. $\epsilon(r, v)$ follows the definition
\begin{align}
    \epsilon(r, v) = -16 \pi^2 G^2 m \rho 
     \int_0^{v_{\rm{esc}}} dv_{\rm{f}} \, v_{\rm{f}}^2 f(v_{\rm{f}}) \mathcal{H}.
\end{align}

Given the power-law density profile of dark matter spike, the distribution function of dark matter particles can be derived following Eddington's formula as
\begin{align}
    f(v_{\rm{f}}) = 
    \frac{\Gamma(\gamma + 1)}{\Gamma\left(\gamma - \frac{1}{2}\right)}
    \frac{1}{2 \gamma \pi^{3/2} v_{\rm{c}}^{2\gamma}}
    \left(2v_{\rm{c}}^2 - v_{\rm{f}}^2\right)^{\gamma - 3/2}.
\end{align}

\subsection{Effect on EMRIs Formation}\label{subsec:formation}
%As discussed above, the relaxation process can significantly affect the orbital eccentricity, allowing the periapsis to approach the SMBH closely enough to emit GWs. The dynamical friction from the dark-matter spike contributes to orbital decay and circularization, resembling the effect of GW emission. 

The total evolution of the inspiraling orbit is determined by the combined effects of dynamical friction and GW emission
\begin{align}
    \left\langle \frac{da}{dt} \right\rangle = 
    \left\langle \frac{da}{dt} \right\rangle_{\rm{df}} + 
    \left\langle \frac{da}{dt} \right\rangle_{\rm{gw}}, \\
    \left\langle \frac{de}{dt} \right\rangle = 
    \left\langle \frac{de}{dt} \right\rangle_{\rm{df}} + 
    \left\langle \frac{de}{dt} \right\rangle_{\rm{gw}}.
\end{align}

The orbit of the test object can be approximated as an elliptical Keplerian orbit. Consequently, the GW emission drives the orbital evolution as \citep{Peters1964}
\begin{align}
\left\langle \frac{da}{dt} \right\rangle_{\rm{gw}} &= 
-\frac{64}{5} \frac{G^3 \mu M^2}{c^5 a^3 (1 - e^2)^{7/2}} 
\left( 1 + \frac{73}{24} e^2 + \frac{37}{96} e^4 \right), \\
\left\langle \frac{de}{dt} \right\rangle_{\rm{gw}} &= 
-\frac{304}{15} \frac{e \, G^3 \mu M^2}{c^5 a^4 (1 - e^2)^{5/2}} 
\left( 1 + \frac{121}{304} e^2 \right).
\end{align}
Here, we consider a binary with component masses $m_1$ and $m_2$, which thus has total mass $M = m_1 + m_2$ and reduced mass $\mu = m_1m_2/M$. We can define the circularization time as
\begin{align}
    t_{\rm{r}} = \frac{1 - e}{d(1 - e)/dt} = -\frac{1 - e}{(de/dt)}.
\end{align}

%The mass of dark matter particles is negligible compared to that of CO. Consequently, dark matter only contributes to dynamical friction without altering the relaxation time $t_{\rm{rlx}}$. For further details, refer to Appendix~\ref{appendix:A}.  

When $t_{\rm{rlx}} < t_{\rm{r}}$, the relaxation process plays a dominant role in the evolution of the system. This leads to an increase in orbital eccentricity, with the semi-major axis remaining approximately constant. Conversely, when $t_{\rm{rlx}} > t_{\rm{r}}$, the evolution of the system is primarily governed by dynamic friction or GW emission, resulting in orbital decay and circularization.

\begin{figure}[t]
    \centering
    \includegraphics[width=1\linewidth]{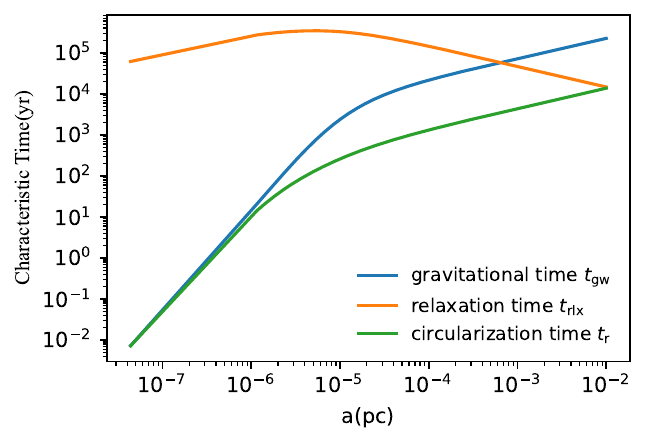}
    \caption{The timescales of various dynamical effects are shown. As the semi-major axis decreases, circularization dominates the orbital evolution. When $a < 0.01\,\rm{pc}$, the relaxation effect becomes negligible.}
    \label{fig:CharacteristicTime}
\end{figure}

It should be noted that the eccentricity of E-EMRIs is initially extremely high, then the effect of dynamical friction, as well as GW emission, generally leads to the gradual circularization of the orbit~\citep{Li:2021pxf, Chen:2025jch}.  On the other hand, field COs contribute to the relaxation effect that drives the growth of orbital eccentricity. The characteristic timescales of these competing effects, under a dark matter density profile with power-law index $\gamma = 3.5$, are shown in Fig.~\ref{fig:CharacteristicTime}, along with the corresponding orbital evolution. As the orbit evolves, we observe that the circularization timescale becomes significantly shorter than the relaxation timescale when $a < 0.01\,\rm{pc}$, indicating that the relaxation effect becomes negligible in this regime.

\begin{figure}[t]
    \centering
    \includegraphics[width=1\linewidth]{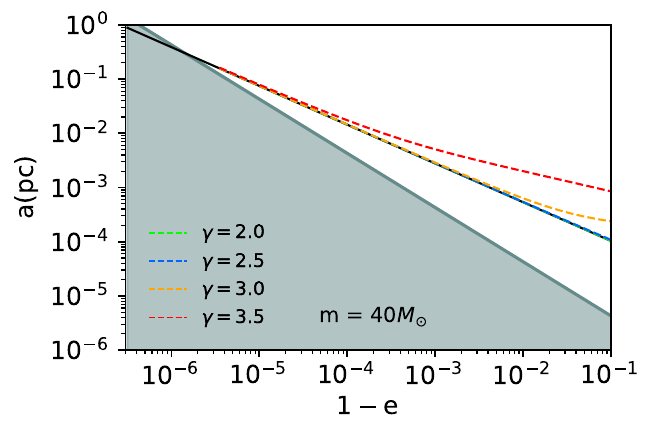}
    \includegraphics[width=1\linewidth]{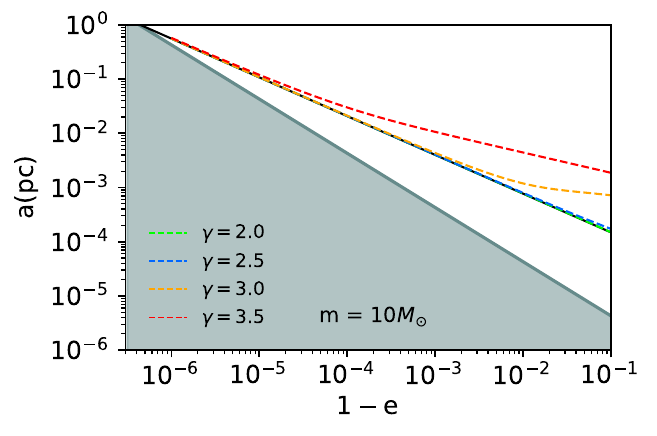}
    \caption{EMRIs formation in the $a-(1-e)$ plane for the inspiral of a black hole with mass $m = 40 M_{\odot}$ (Upper) and $m=10 M_{\odot}$ (Lower) into an SMBH of mass $4.3 \times 10^6 M_{\odot}$. The colored area at bottom is the loss cone region, where no stable orbits exist. The boundary line of this region represents the plunge orbit; Objects in this orbit will be swallowed by the SMBH. The black line represents $t_{\rm{rlx}} = t_{\rm{gw}}$. The green, blue, orange, and red dashed curves represent $t_{\rm{rlx}} = t_{\rm{r}}$ for dark matter spikes with power-law indices of 2.0, 2.5, 3.0, and 3.5, respectively.}
    \label{fig:emri_form_40}
\end{figure}

The impact of dark matter can be seen in Fig.~\ref{fig:emri_form_40}. The region above each curve is dominated by the relaxation process for the respective parameters. Between the loss cone region and the relaxation-dominated region, objects will inspiral into the SMBH. The object that enters this regime will become an EMRI. In the relaxation-dominated region, $1 - e$ will decrease to a very small value, and the object will eventually migrate to the loss cone region or the inspiral region.

The dynamical friction due to dark matter will reduce the extent of the relaxation-dominated region at low eccentricity. The curve is elevated, compared to the pure GW effect. The higher the value of $\gamma$ and the density of dark matter, the more observable the effect becomes.

The intersection between the plunge orbit and each curve representing $t_{\rm{rlx}} = t_{\rm{r}}$ defines the critical semi-major axis $a_{\rm{cri}}$ for each set of dark matter parameters.  Fig.~\ref{fig:emri_form_40} demonstrates that $a_{\rm{cri}}$ remains the same across all cases, suggesting the dynamical friction exerted by dark matter does not influence the event rate, as we shall show next.

\section{Number of sources}\label{sec:number of source}
This section outlines the methodology for estimating the source population. First, we calculate the event rate and subsequently determine the number of sources, while also investigating the influence of dark matter.

Sources with a semi-major axis below $a_{\rm{cri}}$ can form EMRIs. However, to reach the detector's sensitivity band, the source must undergo a significant period of evolution. We denote $a_{\rm{band}}$ as the semi-major axis at which the source enters the detector's sensitivity band. To calculate the number of sources within the band, we can introduce the event rate as~\citep{Hopman_2005}
\begin{align}
    \Gamma = \int_{a_{\rm{min}}}^{a_{\rm{cri}}} \frac{da \, N_{\rm{iso}(a)}}{\ln(L_{\rm{m}} / L_{\rm{lc}}) t_{\rm{rlx}}(a)}.
\end{align}
Here, $L_{\rm{lc}}$ and $L_{\rm{m}}$ represent the loss cone angular momentum and the maximum angular momentum, respectively, while the semi-major axis is $a$. $N_{\rm{iso}}(a)$ represents the number density. The quantity $a_{\rm{min}} = 4GM_{\rm BH}/c^2$, represents the minimum orbital distance at which a stable orbit can exist.

As discussed above, dark matter dynamical friction does not affect $a_{\rm{min}}$, $a_{\rm{cri}}$, or $t_{\rm{rlx}}$. As a result, it does not alter the event rate. Following the discussion in \citep{Amaro_Seoane_2019}, the event rate in Galactic Center can be expressed as \citep{seoane2024eemris}
\begin{multline}
\Gamma \sim 1.92 \times 10^{-6} \, \rm{yrs}^{-1} \tilde{N}_0 \tilde{\Lambda} \tilde{R}_0^{-2} \tilde{m}^2 \\
\times \Biggl\{1.6 \times 10^{-1} \tilde{R}_0^{1/2} \tilde{N}_0^{-1/2} \tilde{\Lambda}^{-1/2} \tilde{m}^{1/2} \mathcal{W}^{-5/4} \\
\times \biggl[ \ln\Bigl( 9138 \tilde{R}_0 \tilde{N}_0^{-1} \tilde{\Lambda}^{-1} \tilde{m} \mathcal{W}^{-5/2} \Bigr) - 2 \biggr] \\
- 4 \times 10^{-2} \tilde{R}_0^{1/2} \times \biggl[ \ln \Bigl( 618 \tilde{R}_0 \Bigr) - 2 \biggr] \Biggr\}.
\end{multline}
Here, the notation is following as
\begin{align*}
    \tilde{N}_0 = \frac{N_0}{12000}, \quad \tilde{R}_0 = \frac{R_{\rm{h}}}{1 \rm{pc}}, \\
    \tilde{m} = \frac{m}{10M_{\odot}}, \quad \tilde{\Lambda} = \frac{\ln \Lambda}{13}.
\end{align*}
The quantity $ N_0 $ represents the number of stellar-mass black holes enclosed within the influence radius $ R_{\rm{h}} $ of the SMBH. Here, $ m $ denotes the mass of a stellar-mass black hole. For the sake of simplicity and illustration we consider the case that all stellar-mass black holes have identical masses. 

We adopt $N_0 = 2 \times 10^{4}$ as the characteristic number of objects in the Galactic Center \citep{Amaro_Seoane_2018}, with an influence radius of $R_{\rm h} = 1~\mathrm{pc}$. Recent studies indicate the existence of multiple formation channels for stellar-mass black holes with masses of $m = 40~\mathrm{M}_{\odot}$ and $10~\mathrm{M}_{\odot}$ \citep{burrows2024}. Consequently, we adopt these two mass values ($m = 40~\mathrm{M}_{\odot}$ and $10~\mathrm{M}_{\odot}$) for stellar-mass black holes in our analysis for illustration and comparison. The Coulomb logarithm is expressed as $\ln\Lambda = \ln(M_{\rm BH}/m)$, where $M_{\rm BH}$ is the mass of SMBH.

To evaluate the number of sources $ N $ within a specific semi-major axis $ a $, we need to consider the line density function \citep{Amaro_Seoane_2019}, given by
\begin{equation}\label{eq:def_line_density}
    g = \frac{dN}{da}.
\end{equation}
We have the current conservation relation
\begin{equation}
    \frac{\partial}{\partial a} \left( \dot{a}(a,e)g \right) + \frac{\partial g}{\partial t} = 0.
\end{equation}
Since the density function remains constant over time, the second term vanishes. After integration, we obtain
\begin{equation}
    \dot{a}(a,e)g = K,
\end{equation}
where $ K $ is a constant. By using Eq.~\ref{eq:def_line_density}, we have the expression for the number of sources as
\begin{equation}\label{eq:source number}
    \frac{dN}{da} = \frac{K}{\dot{a}(a,e)}.
\end{equation}

When the semi-major axis $a$ falls below the critical value $a_{\rm{cri}}$, the system evolves into an EMRI. As the system continues to evolve, once $a$ decreases below the detection threshold $a_{\rm{band}}$, it enters the detector's sensitivity band. The inspiral process terminates when $a$ reaches the minimum stable orbit, $a_{\rm{min}}$, beyond which the CO is swallowed by the SMBH. During the inspiral phase, the CO spends a significant amount of time in eccentric orbits. To characterize the transition to a nearly circular orbit, we define a characteristic semi-major axis $a_{\rm{thr}}$, which marks the critical value at which the orbit becomes circular.

These four characteristic semi-major axes divide the entire evolutionary process into three distinct stages. We denote the corresponding number of sources in these stages as $N_1$, $N_2$, and $N_3$, respectively. These numbers are determined by integrating Eq.~(\ref{eq:source number}), while the population within the observational frequency band can be computed by multiplying the event rate by the time the system spends within the band, $T$. This yields
\begin{equation}
\begin{aligned}
    N_1 &= \int_{a_{\rm{min}}}^{a_{\rm{thr}}} \frac{K}{\dot{a}(a,e)}da, \quad
    N_2 = \int_{a_{\rm{thr}}}^{a_{\rm{band}}} \frac{K}{\dot{a}(a,e)}da,  \\
    N_3 &= \int_{a_{\rm{band}}}^{a_{\rm{cri}}} \frac{K}{\dot{a}(a,e)}da, \quad
    N_1 + N_2 = T \times \Gamma.
\end{aligned}
\end{equation}
Here, $\dot{a}(a,e)$ has been derived in Section~\ref{subsec:formation}. The critical semi-major axis $a_{\rm cri}$ can be determined from Fig.~\ref{fig:emri_form_40}, while $a_{\rm band}$ is defined as the semi-major axis where the signal-to-noise ratio (SNR) reaches 10. In the Galactic center, $a_{\rm min}$ is approximately $8.23 \times 10^{-7} \ \rm{pc}$. The semi-major transition axis $a_{\rm{thr}}$ represents the threshold value at which the second harmonic model becomes dominant. The values of $a_{\rm band}$ and $a_{\rm{thr}}$ vary with different parameters, as expressed in Table~\ref{tab:semi-majoraxis}.

\begin{table}[t]
\centering
\small
\begin{tabular}{@{}|c|c|c|c|c|@{}}
\hline
\multirow{2}{*}{$\gamma$} & 
\multicolumn{2}{c|}{$10\ M_\odot$} & 
\multicolumn{2}{c|}{$40\ M_\odot$} \\
\cline{2-5}
 & $a_{\rm{band}}$ (pc) & $a_{\rm{thr}}$ (pc) & $a_{\rm{band}}$ (pc) & $a_{\rm{thr}}$ (pc) \\ 
\hline
\multicolumn{1}{|c|}{\rm{no DM}} 
& $1.43{\times}10^{-3}$ & $3.43{\times}10^{-6}$ 
& $1.84{\times}10^{-3}$ & $5.69{\times}10^{-6}$ \\
\hline
2   & $1.43{\times}10^{-3}$ & $3.39{\times}10^{-6}$ & $1.84{\times}10^{-3}$ & $5.63{\times}10^{-6}$ \\
\hline
2.5 & $1.43{\times}10^{-3}$ & $3.40{\times}10^{-6}$ & $1.84{\times}10^{-3}$ & $5.42{\times}10^{-6}$ \\
\hline
3   & $1.43{\times}10^{-3}$ & $3.09{\times}10^{-6}$ & $1.84{\times}10^{-3}$ & $4.38{\times}10^{-6}$ \\
\hline
3.5 & $1.43{\times}10^{-3}$ & $2.25{\times}10^{-6}$ & $1.84{\times}10^{-3}$ & $3.23{\times}10^{-6}$ \\
\hline
\end{tabular}
\caption{$a_{\rm{band}}$ and $a_{\rm{thr}}$ for different dark matter profile index $\gamma$ and different stellar-mass black holes.}
\label{tab:semi-majoraxis}
\end{table}

The result of the number of sources in Galactic center is tabulated in Table~\ref{tab:40_number} for different dark matter profile index. As we can observe, when the index $\gamma$ increases, the number of sources decreases steadily. The reason is that although dark matter dynamical friction does not change the event rate, it reduces the total evolution time, resulting in a decrease in the number of sources. 
\begin{table}[t]
\centering
    \begin{tabular}{@{}|c|c|c|c|c|c|@{}}
        \hline
        \multicolumn{2}{|c|}{\textbf{mass of CO}} & \multicolumn{2}{c|}{\textbf{10$M_\odot$}} & \multicolumn{2}{c|}{\textbf{40$M_\odot$}} \\
        \hline
        $\gamma$ & $\rho_{\rm c} \ (M_\odot/\mathrm{pc}^3)$ & $N_1$ & $N_2$ & $N_1$ & $N_2$ \\
        \hline
        \multicolumn{2}{|c|}{without DM} &  0.005 & 2.3 & 0.35 & 133 \\
        \hline
        2 & $1.1{\times}10^8$ & 0.005 & 2.3 & 0.34 & 132 \\
        \hline
        2.5 & $6.5{\times}10^7$ & 0.005 & 2.2 & 0.29 & 127 \\
        \hline
        3 & $3.4{\times}10^7$ & 0.003 & 2 & 0.1 & 87 \\
        \hline
        3.5 & $1.5{\times}10^7$ & 0.0005 & 0.6 & 0.01 & 16 \\ 
        \hline
    \end{tabular}
    \caption{Number of sources for different dark matter profile with power-law index $\gamma$ and $\rho_{\rm{c}}$ denotes the reference density at a distance of $0.01\,\rm{pc}$ from the SMBH.}
    \label{tab:40_number}
\end{table}

\section{GW background from E-EMRIs}\label{sec:background}
Next we shall calculate the GW spectrum of E-EMRIs in the Galactic Center. After introducing GW emission of a single EMRI, we estimate the superposed spectrum from many EMRIs and discuss the effects of dark matter. If the GWs from E-EMRIs cannot be distinguished from other components, they would blend into the overall background and diminish the sensitivity to other sources. Here without losing generality we refer them simply as GW background. However we note that this background is propagating from the Milky Way center, different from the stochastic GW background by other cosmological and astrophysical origins from all directions.
 
\subsection{characteristic strain}\label{subsec:strain}
For an E-EMRI sufficiently far from the SMBH, the compact object's orbit maintains Keplerian characteristics. This orbital configuration enables decomposition of the quadrupole gravitational waveform into harmonic components, where the power radiated through the n-th harmonic follows the relation~\citep{GW1963}%\citep{gravitation} 
\begin{align}
    E_n = \frac{32}{5} \frac{G^4}{c^5} \frac{m_1^2 m_2^2 (m_1 + m_2)}{a^5} g(n, e),
\end{align}
where $g(n,e)$ is a function given by
\begin{multline}
    g(n, e) =  \frac{n^4}{32} \Bigg\{ 
        \bigg[ J_{n-2}(ne) - 2e J_{n-1}(ne) + \frac{2}{n} J_n(ne) \\ 
        + 2e J_{n+1}(ne) - J_{n+2}(ne) \bigg]^2 + \frac{4}{3n^2} \big[J_n(ne)\big]^2  \\
        + (1 - e^2) \bigg[ J_{n-2}(ne) - 2 J_n(ne) + J_{n+2}(ne) \bigg]^2 
        \Bigg\}. 
\end{multline}

The strain amplitude, denoted as $h_0$, quantifies the instantaneous GW amplitude measured at a given time. Persistent GW signals, however, may remain detectable for durations spanning years, permitting cumulative integration over time. For instance, the $n$th harmonic of a compact binary system resides near a frequency $f_n$ for a characteristic timescale $\tau_n \sim f_n / \dot{f}_n$ (or equivalently, $N_{\rm{cycle}} \sim f_n^2 / \dot{f}_n$ cycles) \citep{Finn2000}. This persistent nature enhances the detectability of GW emission at $f_n$, necessitating the substitution of the instantaneous strain amplitude $h_{0,n}$ with the characteristic strain amplitude $h_{{\rm{c}},n}$. The latter inherently accounts for coherent signal integration over observational timescales and serves as the critical parameter in the generalized signal-to-noise ratio calculation. The relationship between $h_{{\rm{c}},n}$ and $h_{0,n}$ is expressed as \citep{Finn2000, Moore_2015, Flanagan1998}
\begin{align}
    h_{{\rm{c}},n}^2 = \left( \frac{f_n^2}{\dot{f}_n} \right) h_{0,n}^2= \frac{1}{(\pi D_{\rm L})^2}  \left( \frac{2G \dot{E}_n}{c^3 \dot{f}_n} \right),
\end{align}
where $D_{\rm L}$ is the luminosity distance to the source. For sources within the Milky Way, this corresponds simply to the physical distance to the source. The term $\dot{f}_n$ represents the rate of change of the $n$-th harmonic frequency and is given by
\begin{align}
    \dot{f}_n = \frac{48n}{5\pi} \frac{(GM_{\rm{c}})^{5/3}}{c^5} (2\pi f_{\rm{orb}})^{11/3} F(e).
\end{align}
Here, the chirp mass $M_{\rm{c}}$ is defined as
\begin{align}
    M_{\rm{c}} = \frac{(m_1 m_2)^{3/5}}{(m_1 + m_2)^{1/5}}.
\end{align}
The function $F(e)$, which accounts for the dependence on orbital eccentricity, is given by \citep{GW1963}
\begin{align}
    F(e)  = \frac{1 + \frac{73}{24} e^2 + \frac{37}{96} e^4}{(1 - e^2)^{7/2}}.
\end{align}

\begin{figure}[t]
    \centering
    \includegraphics[width=1\linewidth]{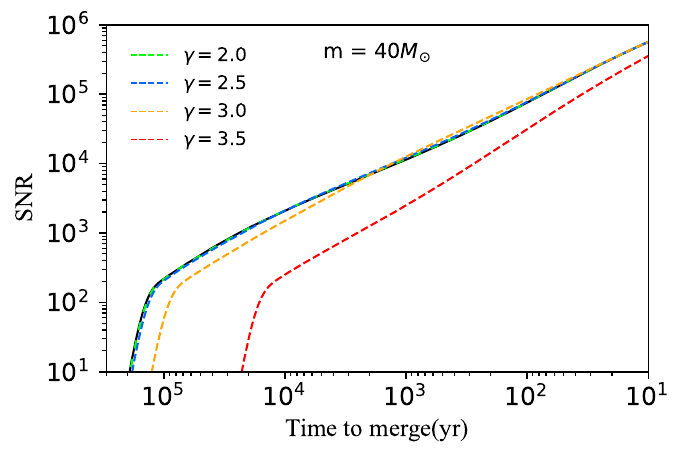}
    \includegraphics[width=1\linewidth]{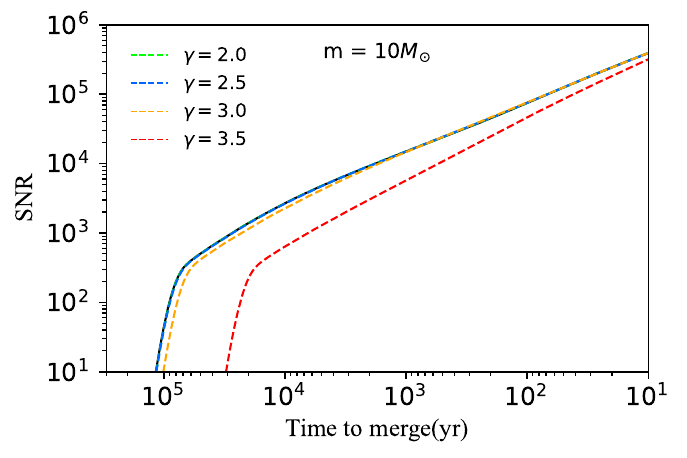}
    \caption{SNR of one-year observation of an E-EMRI, as a function of the remaining time until merger, where the compact object has a mass of $m = 40 M_{\odot}$ (Upper) and $m=10 M_{\odot}$ (Lower). The black curve illustrates the case without dark matter. The dashed curves in green, blue, orange, and red correspond to dark matter power-law indices of 2.0, 2.5, 3.0, and 3.5, respectively.} 
    \label{fig:snr_10}
\end{figure}

Finally, we obtain the expressions for the strain and the characteristic strain \citep{Wagg_2022}
\begin{align}
    h_{{\rm{c}},n}^2 = \frac{2^{5/3}}{3\pi^{4/3}} \frac{(GM_{\rm{c}})^{5/3}}{c^3 D_L^2} \frac{1}{f_{\rm{orb}}^{1/3}} \frac{g(n,e)}{n F(e)}, \\
    h_{0,n}^2 = \frac{2^{28/3}}{5} \frac{(GM_{\rm{c}})^{10/3}}{c^8 D_L^2} \frac{g(n,e)}{n^2} (\pi f_{\rm{orb}})^{4/3}.
\end{align}

The fully averaged SNR can be expressed as \citep{Robson_2019,Wagg_2022}
\begin{align}
    \left\langle \frac{S}{N} \right\rangle^2 = \int_{0}^{\infty} df \frac{h_{\rm{c}}^2}{f^2 S_n(f)}.
\end{align}
Here, $S_n(f)$ represents the power spectral density of the noise. In the general case, a binary system may exhibit eccentricity, requiring a summation over all harmonics. Consequently, the SNR is given by
\begin{align}
     \left\langle \frac{S}{N} \right\rangle^2 = \sum_{n=1}^{\infty} \int_{0}^{\infty} df_n \frac{h_{{\rm{c}},n}^2}{f_n^2 S_n(f_n)}.
\end{align}

For E-EMRIs, orbital evolution is extremely slow, which means that within observation time the frequency undergoes negligible change. Consequently, the orbit can be approximated as stationary, and the SNR can be approximated as
\begin{align}
    \left\langle \frac{S}{N} \right\rangle^2 = \sum_{n=1}^{\infty} \left( \frac{\dot{f}_n}{f_n^2} h_{{\rm{c}},n}^2 \right) \frac{T_{\rm{obs}}}{S_n(f_n)},
\end{align}
where $T_{\rm{obs}}$ denotes the observation time, which will be set to one year for illustration.

We evolved the E-EMRIs orbits and calculated the SNR at each evolutionary moment. For the $40 M_{\odot}$ system, the initial orbital parameters were selected as $a = 0.01$ pc and $e = 0.9995$, whereas for the $10 M_{\odot}$ system, the initial conditions were $a = 0.01$ pc and $e = 0.9997$. The results are presented in Fig.~\ref{fig:snr_10}. When the SNR reaches 10, the semi-major axis of the orbit corresponds to $a_{\rm{band}}$. As listed in Table~\ref{tab:semi-majoraxis}, dark matter does not affect this value. However, as shown in Fig.~\ref{fig:snr_10}, the time to merger at $ \mathrm{SNR} = 10 $ decreases with increasing $\gamma$. This is because dark matter dynamical friction accelerates orbital evolution, thereby shortening the time required for merger. 

\subsection{GW Background}\label{subsec:background}
In the early inspiral phase, E-EMRIs remain at large orbital separations for an extended period due to their slow evolution under gravitational radiation. Given the nearly stationary distribution of sources in the Galactic Center over relevant timescales, the resulting GW background from these systems is expected to be strong and persistent \citep{seoane2024eemris}.

The characteristic strain of GW background from the E-EMRIs, $h_{\rm c,gwb}$, can be calculated using the following expression \citep{phinney2001,Bonetti_2020,Sesana_2008}
\begin{equation}
    \small
    \begin{aligned}
        h_{\rm c,gwb}^2(f) &= \frac{1}{2} \int dz \, d\mathcal{M} \, de 
     \sum_{n} 
    \frac{d^4N}{dz \, d\mathcal{M} \, de \, d\ln f_{\rm{orb}}} 
    \frac{h_{{\rm{c}},n}^2(f)}{f T_{\rm{obs}}},
    \end{aligned}
\end{equation}
where $f_{\rm{orb}} = {f(1+z)}/{n}$, and $n$ denotes the $n$-th harmonic. Since we focus exclusively on the Galactic Center, we can set $z = 0$. 

The computational procedure is as follows. We first evolve an E-EMRI from its initial orbital parameters all the way to merger, recording the trajectory $\{a(t),e(t)\}$. Knowing that the source population is $N_1$ in the semi-major axis interval $[a_{\min},a_{\rm thr}]$ and $N_2$ in $[a_{\rm thr},a_{\rm band}]$, we randomly draw $N_1$ and $N_2$ timestamps from these two intervals, respectively, and treat each selected instant as an independent E-EMRIs. For every sampled time $t_i$ we record the orbital parameters $\{a(t_i),e(t_i)\}$ and their values one year later, $\{a(t_i+\Delta t),e(t_i+\Delta t)\}$ with $\Delta t = 1~\mathrm{yr}$. Using these pairs of parameters, we compute the characteristic strain emitted during each one-year segment. Summing the contributions from all $N_1+N_2$ segments yields the total GW background produced by the unresolved E-EMRIs population \citep{seoane2024eemris}.

We compute the characteristic strain of GWs for two illustrating cases, compact objects with a mass of $40 M_{\odot}$ or with $10 M_{\odot}$. For the $40 M_{\odot}$ case, the initial orbital parameters were chosen as $a = 0.01$ pc and $e = 0.9995$, while for the $10 M_{\odot}$ case, the initial conditions were set to $a = 0.01$ pc and $e = 0.9997$.

 \begin{figure}[t]
    \centering
    \includegraphics[width=1\linewidth]{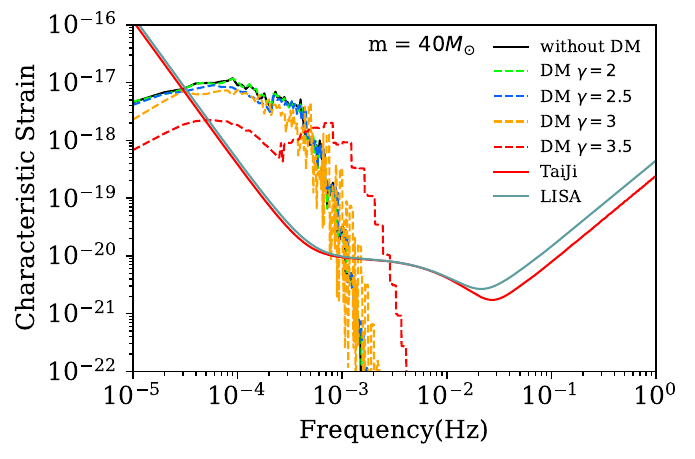}
    \includegraphics[width=1\linewidth]{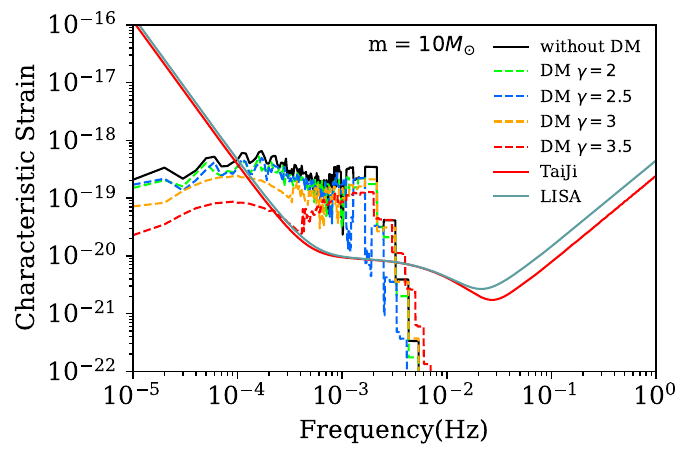}
    \caption{The GW  background of E-EMRIs at the Galactic Center, where the compact objects have a mass of $m = 40 M_{\odot}$ (Upper) and $m = 10 M_{\odot}$ (Lower). The black solid curve represents the case without dark matter, while the green, blue, orange, and red dashed curves for dark matter spikes with power-law indices of 2.0, 2.5, 3.0, and 3.5, respectively. The red and bluish-green solid curves denote the sensitivities of Taiji and LISA with the galactic confusion noise\citep{Liu_2023}.}
    \label{fig:gwb10}
\end{figure}

The LISA/Taiji frequency band is partitioned into bins of width $\delta f \sim 10^{-5}$ Hz. For each E-EMRIs in the sample, we evaluate the characteristic strain contributions from all harmonics within the LISA/Taij band ($10^{-5}$ Hz $< n f_{\rm{orb}} < 10^{-1}$ Hz) and allocate them to the corresponding frequency bins. The contribution of the $n$th harmonic from the $m$th E-EMRIs , represented as $(h_{\rm c,gwb}^2)_{m,n}$, is given by \citep{Bonetti_2020}
\begin{align}
    (h_{\rm c,gwb}^2)_{m,n} =
    \begin{cases}
        \frac{h_{{\rm{c}},n}^2}{2 f T_{\rm{obs}}}, & {\rm{if}} \ \ \ \ f \leq \dot{f} T_{\rm{obs}},
        \\
        \frac{h_n^2}{2}, & {\rm{if}} \ \ \ \ f > \dot{f} T_{\rm{obs}}.
    \end{cases}
\end{align}
Here, $f = n f_{\rm{orb},m}$ denotes the frequency of the $n$th harmonic, while $\Delta f_{m,n}$ represents its frequency variation over the observation period, which is set to $T_{\rm{obs}} = 1$ yr. When $\Delta f_{m,n}$ exceeds the bin width $\delta f$, the harmonic extends across multiple frequency bins, distributing its contribution accordingly. In contrast, if $\Delta f_{m,n} \leq \delta f$, the harmonic remains localized within a single bin.

%  \begin{figure}[t]
%     \centering
%     \includegraphics[width=1\linewidth]{hcgwb40.pdf}
%     \includegraphics[width=1\linewidth]{hcgwb10.pdf}
%     \caption{The GW  background of E-EMRIs at the Galactic Center, where the compact objects have a mass of $m = 40 M_{\odot}$ (Upper) and $m = 10 M_{\odot}$ (Lower). The black solid curve represents the case without dark matter, while the green, blue, orange, and red dashed curves for dark matter spikes with power-law indices of 2.0, 2.5, 3.0, and 3.5, respectively. The red and bluish-green solid curves denote the sensitivities of Taiji and LISA \citep{Liu_2023}.}
%     \label{fig:gwb10}
% \end{figure}

In Fig.~\ref{fig:gwb10} we show the calculated GW background with and without dark matter spikes in two panels. Although the overall characteristic strain amplitude of systems with $m = 10\, M_\odot$ is much lower than that with $m = 40\, M_\odot$, which can be primarily attributed to the mass difference, both cases share similar behaviors. The spectrum shows that as the dark matter power-law index $\gamma$ increases, the impact of dark matter dynamical friction on the GW background of E-EMRIs becomes more pronounced. In particular, at lower frequencies, dynamical friction suppresses the GW background, while at higher frequencies, it may enhances the power spectrum. This occurs because dark matter-induced drag enhances orbital circularization and accelerates the inspiral, reducing the number of sources that remain in the low-frequency regime while circularizing the orbits in the higher frequency regime. Consequently, fewer systems contribute to the background in the low-frequency regime, while a more significant signal is observed at higher frequencies.

\section{CONCLUSIONS}\label{sec:conclusion}
In this work, we have investigated the formation of EMRIs driven by relaxation in the Galactic Center, with a focus on the impact of dark matter-induced dynamical friction. We have modeled the orbital evolution of COs under various dark matter profiles, and computed the resulting EMRIs population as well as the GW background generated by unresolved sources. Our results indicate that dynamical friction from dark matter accelerates orbital evolution, thereby reducing the number of EMRIs and altering the GW background. Specifically, dynamical friction suppresses the GW background  at lower frequencies, but enhances the spectrum at higher frequencies. 
% These findings suggest a potential method for probing the distribution and properties of dark matter near SMBH through future space-based GW observations.

We have shown that E-EMRIs might have high SNR in space-based detectors, such as LISA and Taiji, and the imprint of dark-matter dynamical friction on their waveforms might be sizable. When strong signals are identified, the next step would be performing parameter estimation, and extracting the information of dark-matter spike, which we shall pursue in future work. Also GWs from E-EMRIs in the Galaxy Center have a particular sky location. This directional information~\citep{Guo:2023lzb, Guo:2024zue} may enable us to distinguish them from other stochastic GW background.

\begin{acknowledgments}
This work is partly supported by the National Key Research and Development Program of China (Grant No.2021YFC2201901), 
the National Natural Science Foundation of China (Grant No.12147103),
and the Fundamental Research Funds for the Central Universities. 
\end{acknowledgments}

%% To help institutions obtain information on the effectiveness of their 
%% telescopes the AAS Journals has created a group of keywords for telescope 
%% facilities.
%
%% Following the acknowledgments section, use the following syntax and the
%% \facility{} or \facilities{} macros to list the keywords of facilities used 
%% in the research for the paper.  Each keyword is check against the master 
%% list during copy editing.  Individual instruments can be provided in 
%% parentheses, after the keyword, but they are not verified.

\vspace{5mm}

\bibliography{refs}{}

%% This command is needed to show the entire author+affiliation list when
%% the collaboration and author truncation commands are used.  It has to
%% go at the end of the manuscript.
%\allauthors

%% Include this line if you are using the \added, \replaced, \deleted
%% commands to see a summary list of all changes at the end of the article.
%\listofchanges

\end{document}